\documentclass{article}
\usepackage{amssymb, amsmath}

\begin{document}

\begin{center}
{\Large \ Cyclic codes over some special rings} 
\begin{equation*}
\end{equation*}
Cristina Flaut\bigskip

Faculty of Mathematics and Computer Science,

Ovidius University,

Bd. Mamaia 124, 900527, Constanta,

ROMANIA

e-mail: cflaut@univ-ovidius.ro

cristina\_flaut@yahoo.com

http://cristinaflaut.wikispaces.com/ 
\begin{equation*}
\end{equation*}
\end{center}

\bigskip \textbf{Abstract.} {\small In this paper we will study cyclic codes
over some special rings:\ }$\mathbb{F}_{q}[u]/(u^{i}),\mathbb{F}%
_{q}[u_{1},...,u_{i}]/(u_{1}^{2},u_{2}^{2},...,u_{i}^{2},u_{1}u_{2}-u_{2}u_{1},...,u_{k}u_{j}-u_{j}u_{k},...), 
$\newline
$\mathbb{F}_{q}[u,v]/\left( u^{i},v^{j},uv-vu\right) ${\small , }$q=p^{r}$%
{\small , where }$p${\small \ is a prime number, }$r\in \mathbb{N}-\{0\}$%
{\small \ and \ }$\mathbb{F}_{q}${\small \ is a field with }$q${\small \
elements.\medskip }

\textbf{Key words.} {\small Cyclic codes; Codes over rings; Hamming
distance.\medskip }

\textbf{AMS classification (2000).} {\small 94B15, 94B05.} 
\begin{equation*}
\end{equation*}

{\large 0. Introduction\bigskip }

Codes over finite rings have been intensively studied in the last time, some
of the earliest results of them are in [Bl; 72], [Sp; 78]. Ones of the most
important finite rings in the coding theory are: the finite field $\mathbb{F}%
_{q}$ and the ring $\mathbb{Z}_{q},$ where{\small \ }$p${\small \ }is{\small %
\ }a prime number{\small , }$q=p^{r}${\small , }$r\in \mathbb{N}-\{0\}$. For
example, in the paper [Ha et all; 94] some codes over $\mathbb{Z}_{4}$ are
investigated.

The class of cyclic codes is an important class of linear codes with a big
interest in coding theory. Described as ideals in certain polynomial rings,
they have a good algebraic structure and the cyclic codes over some special
finite rings was recently described (see [Gr; 97], [Ne, Ho; 99], [Qi, Zh,
Zhu; 05], [Ab, Si; 07], [Al-As, Ha; 10], [Yi, Ka; 11], [Do, Ka, Yi; 11] ).
Two classes of \ these main rings are: Galois rings and rings of the form 
{\small \ }$\mathbb{F}_{q}[u]/(u^{i})$ or generalization of these, where $%
q=p^{r}${\small , } with $p$\ a prime number,{\small \ }$r\in \mathbb{N}%
-\{0\}.$

In this paper, we will investigate the structure of cyclic codes of
arbitrary length over the rings:

$\mathbb{F}_{q}[u]/(u^{i}),$

$\mathbb{F}%
_{q}[u_{1},...,u_{i}]/(u_{1}^{2},u_{2}^{2},...,u_{i}^{2},u_{1}u_{2}-u_{2}u_{1},...,u_{k}u_{j}-u_{j}u_{k},...), 
$

$\mathbb{F}_{q}[u,v]/\left( u^{i},v^{j},uv-vu\right) $.

\begin{equation*}
\end{equation*}

{\large 1. Preliminaries}%
\begin{equation*}
\end{equation*}

The \textit{Galois ring} $GR\left( q,n\right) $ is the residue class ring $%
\mathbb{Z}/p^{r}\mathbb{Z[}x\mathbb{]}\ /\ (f(x)),$ where $f\left( x\right) $
is a monic irreducible polynomial of degree $n$ in $\mathbb{Z}_{p^{r}}[x]$
such that $f\left( x\right) $\textit{mod}$p$ is a monic irreducible
polynomial in \ $\mathbb{Z}_{p}[x].$ The existence of the polynomial $%
f\left( x\right) $ is given by the Hensel lifting, which allows us to "lift"
a root $\rho $ of a polynomial $f$\textit{mod}$p^{t}$ to a new root $\sigma $
for the polynomial $f$\textit{mod} $p^{t+1},t\in \mathbb{N}-\{0\}$ (see [Ei;
95], [Mi; 80] ). From here, it results that we can choose \ the polynomial $%
f,$ \ monic and irreducible over $\mathbb{Z}_{p}$, as in the standard
construction of the Galois field $\mathbb{F}_{p^{r}}$ from $\mathbb{Z}_{p},$
and we lift it to a polynomial over $\mathbb{Z}/p^{r}\mathbb{Z}.$ We remark
that $\left\vert GR\left( q,n\right) \right\vert =p^{rm}.$ For example, $%
GR\left( q,1\right) =$ $\mathbb{Z}_{q}$ and $GR\left( p,r\right) =\mathbb{F}%
_{q}.$ Let $\theta $ be a root of the polynomial $f\left( x\right) .$ Since
we can think at $GR\left( q,n\right) $ as a Galois extension $\mathbb{Z}%
_{p^{r}}[\theta ]$ of $\mathbb{Z}_{p^{r}}$ by a root $\theta $ of $f\left(
x\right) ,$ therefore each element $v\in GR\left( q,n\right) $ has the form 
\begin{equation*}
v=a_{0}+a_{1}\theta +a_{2}\theta ^{2}+...+a_{n-1}\theta ^{n-1},
\end{equation*}%
where $a_{i}\in \mathbb{Z}_{p^{r}},i\in \{0,1,...,n-1\}$ (see [So, Si; 07]
and [So, Si; 07(1)]).

The Galois ring $GR\left( q,n\right) $ is a free module of rank $n$ over $%
\mathbb{Z}_{q}$ and the set 
\begin{equation*}
\{1,\theta ,\theta ^{2},...,\theta ^{n-1}\}
\end{equation*}%
is a free basis for $GR\left( q,n\right) .$ Since the ring $\mathbb{Z}%
_{p^{r}}$ satisfies the invariant dimension property, it results that all
bases in $GR\left( q,n\right) $ have $n$ elements.

Let $R$ \ be a commutative ring and $I$ be an ideal of the ring $R.$ The
ideal \ $I$ is called \textit{principal} \ if it is generated by one
element. The ring $R$ is called \textit{principal} if all its ideals are
principal. The ring $R$ is called a \textit{local ring} if it has a unique
maximal ideal. A ring $R$ is called \textit{chain ring } if the set of all
ideals of $R$ is ordered by inclusion (is a chain under set inclusion). For
a chain ring its unique maximal ideal contains the nilpotent elements.

All ideals in a finite chain ring \ $R$ are principal. Indeed, if $I$ is not
\ a principal ideal, since $R$ is finite, we have that $I$ is finite
generated and $I=<a_{1},...,a_{t}>,$ where $\{a_{1},...,a_{t}\}$ is a
minimal set of generators. It results $<a_{i}>\varsubsetneq <a_{j}>,i\neq
j,i,j\in \{1,2\},$ which is a contradiction, since $R$ is a chain ring. We
obtain that all ideals in a finite chain ring are principal and there is a
unique maximal \ ideal. It results that a chain ring is a local ring. For
details about the chain rings, the reader is referred to [McD; 74].

Let $\boldsymbol{m}$ be the maximal ideal in a finite chain ring and let $u$
be its generator, i.e. $\boldsymbol{m}=<u>=Ru.$ Since $R$ is finite, the
chain $R=<u^{0}>\supseteq <u^{1}>\supseteq <u^{2}>\supseteq
....<u^{j}>\supseteq ....$ is a finite chain. It results that there is an
element $j$ with the property $<u^{j}>=0.$ The smallest number $t$ such that 
$<u^{t}>=0$ is called $\ $\textit{the nilpotency index }of $u.$ The residue
field $\mathbb{F}=R/\boldsymbol{m}$ has $q=p^{t}$ elements with $p$ a prime
number, $char\mathbb{F}=p$ and $\left\vert \mathbb{F}^{\ast }\right\vert
=p^{t}-1.$

For details about the finite chain rings, the reader is referred to [McD;
74].

Galois rings or the rings of the form {\small \ }$\mathbb{F}_{q}[u]/(u^{i})$
are principal ideal rings.

Galois ring $GR\left( p^{r},n\right) $ is a finite chain ring (of length $%
r). $

Finite chain rings allow us to find good description for cyclic codes over
these rings.

Let $R$ be a unitary finite commutative ring. A \textit{code} $C$ of length $%
n$ over $R$ is a nonempty subset of $R^{n}=\underset{n-times}{\underbrace{%
R\times R\times ...\times R}}$. The elements of $C$ are called \textit{%
codewords}. A \textit{linear code} $C$ of length $n$ over $R$ is a $R$%
-submodule of $R^{n}$. We remark that such a submodule is not necessary a
free module. A linear code $C$ of length $n$ is a \textit{cyclic code} if
for each codeword $c=(c_{0},...,c_{n-1})\in C,$ the codeword $\left(
c_{n-1},c_{0},...,c_{n-2}\right) $ belongs to $C.$

For the cyclic codes, we will write the codewords as polynomials. Let $C$ be
a cyclic code. For \ each $c=(c_{0},...,c_{n-1})\in C$ we associate the
polynomial $c\left( x\right) $ of degree less than $n,$ $c\left( x\right)
=c_{0}+c_{1}x+...+c_{i}x^{i}+...+c_{n-1}x^{n-1}\in R[x],$ called \textit{the
associated polynomial}. The codeword $\overline{c}=\left(
c_{n-1},c_{0},...,c_{n-2}\right) \ \ $has the associated polynomial $\ 
\overline{c}\left( x\right)
=c_{n-1}+c_{0}x+...+c_{i}x^{i+1}+...+c_{n-2}x^{n-1}$ and we have $\overline{c%
}\left( x\right) =c\left( x\right) x-c_{n-1}(x^{n}-1),$ therefore $\overline{%
c}\left( x\right) =c\left( x\right) x$ \textit{mod} $\left( x^{n}-1\right) $%
. We remark that $c\left( x\right) \in C\ $\textit{mod} $\left(
x^{n}-1\right) $ if and only if $c\left( x\right) x\in C$ \textit{mod} $%
\left( x^{n}-1\right) $. Using induction steps, $c\left( x\right) x\in C\ $%
\textit{mod} $\left( x^{n}-1\right) $ if and only if $c\left( x\right)
x^{2}\in C$\textit{mod} $\left( x^{n}-1\right) .$ Therefore we have $c\left(
x\right) x^{i}\in C$ \textit{mod} $\left( x^{n}-1\right) ,$ for all $i\in 
\mathbb{N}-\{0\}.$ From here, it results that $C$ is a cyclic code of length 
$n$ over $R$ if and only if $C$ is an ideal in the ring $R[x]/\left(
x^{n}-1\right) .\bigskip $

{\large 2.~The\ rings\bigskip }

With the above notations, we consider the rings 
\begin{equation*}
R_{i}\simeq \mathbb{F}_{q}[u]/(u^{i}),
\end{equation*}%
\begin{equation*}
S_{i}\simeq \mathbb{F}%
_{q}[u_{1},...,u_{i}]/(u_{1}^{2},u_{2}^{2},...,u_{i}^{2},u_{1}u_{2}-u_{2}u_{1},...,u_{k}u_{j}-u_{j}u_{k},...),k\neq j,
\end{equation*}%
\begin{equation*}
T_{(i,j)}=\mathbb{F}_{q}[u,v]/\left( u^{i},v^{j},uv-vu\right) ,i,j\in 
\mathbb{N}-\{0\}.
\end{equation*}

For example, the ring$\ \ \ R_{i}$ is a commutative chain ring and $<u>$ is
a maximal ideal\ (see \textbf{\ }[Do, Li, Pa; 11])\medskip .

For $R\in \{R_{i},S_{i},T_{(i,j)}\},$ $i,j,n\in \mathbb{N}-\{0\},$we denote

\begin{equation*}
R_{i,n}=R_{i}[x]/\left( x^{n}-1\right) ,
\end{equation*}%
\begin{equation*}
S_{i,n}=S_{i}[x]/\left( x^{n}-1\right) ,
\end{equation*}%
\begin{equation*}
T_{(i,j),n}=T_{(i,j)}[x]/\left( x^{n}-1\right) .
\end{equation*}

\textbf{Remark 2.1.} Since the rings $R_{i},S_{i},T_{(i,j)}$ are finite
rings, the rings $R_{i,n},$\newline
$S_{i,n},T_{(i,j),n}$ are isomorphic with the group ring $RG,$where $G=(g\
/\ g^{n}-1=0)$ is the cyclic group of order $n$ and$\ \ R\in
\{R_{i,n},S_{i,n},T_{(i,j),n}\}.\medskip $

\textbf{Remark 2.2.} If the characteristic of the ring is not prime with the
length $n$ of the code$,$ then the polynomial $x^{n}-1$ factors uniquely
over $\mathbb{F}_{q}$, but does not factor uniquely over the rings $%
R_{i},S_{i},T_{(i,j)}$. Indeed, for example, for $p=2,r=2,i=3,j=2,n=2,$ we
have $x^{2}-1=(x-1)^{2}=(x-(1-u^{2}))^{2}$ over $R_{3},$ $%
x^{2}-1=(x-1)^{2}=(x-(1+u_{1}^{2}+u_{2}^{2}+u_{3}^{2}))^{2}$ over $S_{3}$, $%
x^{2}-1=(x-1)^{2}=\left( x-\left( 1+u^{2}+v\right) \right) ^{2}$ over $%
T_{(3,2)}.$

In [Qi, Zh, Zhu; 05], if $\gcd \left( n,p\right) =1,$ the authors proved
that $R_{i,n}$ is a principal \ ideal ring. In the case of the rings $%
S_{i,n} $ and \ $T_{(i,j),n}$ \ situation is not the same.\medskip\ 

\textbf{Proposition \ 2.3.} \textit{The rings} \ $S_{i,n}$ \textit{and} \ $%
T_{(i,j),n}\ \ $\textit{are not principal ideal rings.\medskip }

\textbf{Proof.} In the following, we will use some ideas given in [Yi, Ka;
11], Lemma 2.4, when the authors proved the above result for $S_{i,n}$ in
the case when $i=2,p=2,r=1.$

Let $R\in \{S_{i,n},T_{(i,j),n}\}.~$From Remark 2.1, we define the ring
morphism $\varphi :RG\rightarrow R,\varphi \left(
c_{0}+c_{1}x+...+c_{n-1}x^{n-1}\right) =c_{0}+c_{1}+...+c_{n-1}$, which is a
surjective map, called the \textit{augmentation morphism}. Let $%
I_{i,n}=(u_{1},...,u_{i})$ be the ideal in $S_{i,n}$ generated by the
elements $\{u_{1},...,u_{i}\}$ and $I_{u,v}$ be the ideal in $T_{(i,j),n}$
generated by the elements $\{u,v\}.$ These ideals are not principal ideals$.$
Let \ $I\in \{I_{i,n},I_{u,v}\}.$ We have that $\varphi ^{-1}\left( I\right)
=J$ \ is an ideal in $R.$ Since $\varphi $ is surjective, therefore \ $%
\varphi \left( J\right) $ is an ideal in $R$ and $\varphi \left( J\right) =$ 
$\varphi \left( \varphi ^{-1}\left( I\right) \right) =I$. From here, if $\ J$
is a principal ideal, it results that $I~\ $is a principal ideal, false.$%
~\Box \medskip $

\textbf{Proposition 2.4.} \textit{With the above notations,} \textit{for }$%
n=p^{l}k,$ \textit{with} $k>1,$\newline
$\gcd (p,k)=1,$ \textit{the rings} \ $R_{i,n},S_{i,n}$ \ \textit{and} \ $%
T_{(i,j),n}\ $ \textit{are not local rings}.\medskip

\textbf{Proof.} From [Hu; 74], we know that a ring is local if and only if
the non-units form a maximal ideal in the ring. Let $R\in
\{R_{i,n},S_{i,n},T_{(i,j),n}\}.$

\textit{Case 1.} $\gcd (p,k-1)\neq 1.$ \ From hypothesis, in $R$, we have 
\begin{equation*}
0=x^{p^{l}k}-1=(x^{p^{l}}-1)\left(
x^{p^{l}(k-1)}+x^{p^{l}(k-2)}+...+1\right) .
\end{equation*}%
It results that $f\left( x\right) =$ $x^{p^{l}(k-1)}+x^{p^{l}(k-2)}+...+1$
is a zero divisor, then it is not invertible in $R.$ We obtain $\varphi
\left( x^{p^{l}(k-2)}+x^{p^{l}(k-2)}+...+1\right) =\underset{(k-1)-times}{%
\underbrace{1+1+...+1}}=0.$ Then $g\left( x\right)
=x^{p^{l}(k-2)}+x^{p^{l}(k-2)}+...+1$ is a non-unit in $R$. Since $%
x^{p^{l}(k-1)}x^{p^{l}}=1,$we have that $f\left( x\right) -g\left( x\right)
=x^{p^{l}(k-1)}$ is a unit in \ $R.$ Therefore non-invertible elements in $R$
do not form an ideal, hence $R$ is not a local ring.

\textit{Case 2.} \ $\gcd (p,k-1)=1.$ Let $g_{1}=ug,$ $g_{1}$ be a non-unit
element. Such an element is a nilpotent element. $\ $Denoting $h\left(
x\right) =f\left( x\right) +g_{1}\left( x\right) ,$ we have $\varphi \left(
h\left( x\right) \right) =k+\left( k-1\right) u.$ Since $\gcd (p,k-1)=1,$ it
results that $\left( k-1\right) u$ is a nonzero nilpotent element. From
here, we obtain that $k+\left( k-1\right) u$ is a sum between an invertible
element, $k,$ and a nilpotent element, therefore it is an invertible
element. It results that $\varphi \left( h\left( x\right) \right) $ is a
unit and $h\left( x\right) $ is also a unit, hence non-invertible elements
do not form an ideal. We just proved that $R$ is not a local ring.$~\Box
\medskip $

\textbf{Proposition 2.5.} \ \textit{With the above notations,} \textit{for }$%
n=p^{l}$ \ \textit{the rings} \ \ $R_{i,n},$\ $S_{i,n},T_{(i,j),n}\ $ 
\textit{are local rings}.\medskip

\textbf{Proof.} Let $R_{n}\in \{R_{i,n},S_{i,n},T_{(i,j),n}\},R\in
\{R_{i},S_{i},T_{(i,j)}\}.$ We will prove that the non-units form an ideal
in the ring \ $R_{n}$. First of all, we \ remark that a non-unit element in $%
R_{n}$ different from zero is a zero divisor. Indeed, if $\alpha \in
R_{n},\alpha \neq 0,$ is a non-invertible element, then the ideal generated
by $\alpha $ is different from $R_{n}$. Hence we can find the elements $%
\alpha _{1}\neq \alpha _{2}$ such that $\alpha _{1}\alpha =\alpha _{2}\alpha
.$ Therefore $\left( \alpha _{1}-\alpha _{2}\right) \alpha =0$ and $\alpha $
is a zero divisor. We remark that an element~ $\theta $ ~in\ \ $R$ has the
form $\theta =\theta _{1}+\theta _{2},$ where $\theta _{1}\in \mathbb{F}_{q}$
and\ \ $\theta _{2}\in R-$ $\mathbb{F}_{q}$.\ \ \ If \ $\theta $ is a unit
in $R$, then $\theta _{1}\in \mathbb{F}_{q}^{\ast }$. Let \ $%
s=s_{0}+s_{1}x+...+s_{n-1}x^{n-1}\in R_{n}$ \ and $%
s_{k}=(s_{k})_{1}+(s_{k})_{2},$ where $(s_{k})_{1}\in \mathbb{F}_{q}\ $ and\ 
$(s_{k})_{2}\in R-\mathbb{F}_{q},j\in \{1,...,n-1\}$ $.$ \ Since $%
x^{p^{j}}=1,$ it results that $%
s^{p^{j}}=s_{0}^{p^{j}}+s_{1}^{p^{j}}+...+s_{n-1}^{p^{j}}$ and \ $s^{p^{j}}$
is a unit or a zero divisor. If $\ s^{p^{j}}\ $is a zero divisor, then$\ \ s$
~is a zero divisor, hence a non-unit. If \ $s^{p^{j}}$ is a unit, it results
that $s$ is a unit, since $s\cdot s^{p^{j}-1}=$\ $s^{p^{j}}$. \ If \ $%
s^{p^{j}}$ is a zero divisor, then $\sum%
\limits_{k=1}^{n-1}(s_{k}^{p^{j}})_{1}=0$ and this characterize a
zero-divisor, hence a non-unit in the ring $R_{n}.$ Let \ \ $%
t=t_{0}+t_{1}x+...+t_{n-1}x^{n-1}\in R_{n},$ with \ $\sum%
\limits_{k=1}^{n-1}(t_{k}^{p^{j}})_{1}=0,$ be another non-unit. The element $%
r=s+t$ \ is also a non-unit. For prove this, we compute $\sum%
\limits_{k=1}^{n-1}(r_{k}^{p^{j}})_{1}.$It results $\sum%
\limits_{k=1}^{n-1}(r_{k}^{p^{j}})_{1}=\sum%
\limits_{k=1}^{n-1}(s_{k}+t_{k})_{1}^{p^{j}}=\sum%
\limits_{k=1}^{n-1}(s_{k}^{p^{j}})_{1}+\sum%
\limits_{k=1}^{n-1}(t_{k}^{p^{j}})_{1}=0+0=0,$ therefore $r$ is a non-unit
and non-units form an ideal. Hence $R$ is a local ring.\textbf{\ }$\Box
\medskip $

Propositions 2.5 and 2.6 was proved for the ring \ $S_{i,n}$ in \ [Yi, Ka;
11] in the particular case $i=2$ ( see Theorem 2.5 and Theorem 2.7).

\begin{equation*}
\end{equation*}

{\large 3. Ranks \ for the cyclic codes over the rings \ }$%
R_{i},S_{i},T_{(i,j)}\bigskip \bigskip $

In [Bh, Wa; 09], Proposition 1, the authors described cyclic codes of length 
$n$ over the Galois ring $GR\left( q,l\right) =\mathbb{Z}_{p^{m}}[x]/\left(
f\right) ,\deg f=l,q=p^{m},\left( n,q\right) =1.$ Using some ideas given in
this proof, we can describe cyclic codes over the rings $%
R_{i},S_{i},T_{(i,j)}$ in the general case.\medskip

\textbf{Proposition 3.1.} \textit{Let} $\ R\in \{R_{i},S_{i},T_{(i,j)}\}$. 
\textit{A non-zero cyclic code} $C$ \textit{of length} $n$ \textit{over} $R$ 
\textit{is a free module over} $R$ \textit{if it is generated by a monic
polynomial} $h\left( x\right) ,h\left( x\right) \mid \left( x^{n}-1\right) $ 
\textit{over} $R.$ \textit{In this case,} $rankC=n-r,~\deg h\left( x\right)
=r,$ \textit{and} $\{h\left( x\right) ,xh\left( x\right)
,...,x^{n-r-1}h\left( x\right) \}$ \textit{is a basis in} $C.\medskip $

\textbf{Proof.} Let $C$ be a non-zero cyclic code $C$ of length $n$ over $R$
generated by the polynomial$\ h\left( x\right) ,h\left( x\right) \mid \left(
x^{n}-1\right) .$ Since $h\left( x\right) \mid \left( x^{n}-1\right) ,$we
can consider $h\left( x\right) $ a monic polynomial. Then there is a monic
polynomial $q\left( x\right) \in R$ such that $h\left( x\right) q\left(
x\right) =x^{n}-1.$ From here, it results that $\deg q\left( x\right) =n-r$
and $\ q$ has the form $q\left( x\right) =q_{0}+q_{1}x+...+x^{n-r}.~$Let $%
R_{n}\in \{R_{i,n},S_{i,n},T_{(i,j),n}\}.$ In~ $R_{n}$ we have $h\left(
x\right) q\left( x\right) =h\left( x\right) (q_{0}+q_{1}x+...+x^{n-r})=0.$
Therefore $x^{i}h\left( x\right) ,$ for $i\geq n-r,$ can be written as a
linear combination of the elements $\{h\left( x\right) ,xh\left( x\right)
,...,x^{n-r}h\left( x\right) \},$ hence each element in $C$ \ of the form $%
p\left( x\right) h\left( x\right) ,p\left( x\right) \in R_{n}$ is a linear
combination of $\{h\left( x\right) ,xh\left( x\right) ,...,x^{n-r-1}h\left(
x\right) \}.$ It results that the system\newline
$\{h\left( x\right) ,xh\left( x\right) ,...,x^{n-r-1}h\left( x\right) \}$
spans $C$. For linearly independence over $R$, let $\alpha _{0},...,\alpha
_{n-r-1}\in R$ such that $\alpha _{0}h\left( x\right) +\alpha _{1}xh\left(
x\right) +...+\alpha _{n-r-1}x^{n-r-1}h\left( x\right) =0.$ We obtain $%
(\alpha _{0}+\alpha _{1}x+...+\alpha _{n-r-1}x^{n-r-1})h\left( x\right) =0$
in $R_{n},$ hence $\left( x^{n}-1\right) \mid (\alpha _{0}+\alpha
_{1}x+...+\alpha _{n-r-1}x^{n-r-1})h\left( x\right) $ in $R[x]$, with 
\begin{equation*}
\deg (\alpha _{0}+\alpha _{1}x+...+\alpha _{n-r-1}x^{n-r-1})h\left( x\right)
=n-1<n.
\end{equation*}%
From here, we have $(\alpha _{0}+\alpha _{1}x+...+\alpha
_{n-r-1}x^{n-r-1})h\left( x\right) =0$ in $R[x].$ Since $h\left( x\right) $
is monic, it results $\alpha _{0}+\alpha _{1}x+...+\alpha
_{n-r-1}x^{n-r-1}=0,~$hence $\alpha _{0}=\alpha _{1}=...=\alpha
_{n-r-1}=0.~\Box \smallskip $

We remark that another proof of the above results can be obtained \ using
the main result from [Gr; 97].\medskip

\textbf{Corollary 3.2.} \textit{With the notations used in Proposition 3.1,
if }$C$\textit{\ is a nonzero-cyclic code} \textit{of length} $n$ \textit{%
over} $R$ \textit{generated by a monic polynomial} $h\left( x\right) ,$%
\newline
$h\left( x\right) \mid \left( x^{n}-1\right) $ \textit{over} $R,$ \textit{%
then} \ $\left\vert C\right\vert =\left\vert R\right\vert ^{n-r}.~\Box
\medskip $

\textbf{Proposition 3.3.} \textit{Let} $\ $ $C$ \textit{be a} \textit{%
non-zero cyclic code of length} $n$ \textit{over} $R\in
\{R_{i},S_{i},T_{(i,j)}\}$ \textit{\ generated by the polynomials }$%
\{h_{1},...,h_{t}\}.$ \textit{Therefore} $C$ \textit{is a \ vector space over%
} $\mathbb{F}_{q}$ \textit{and }$\left\vert C\right\vert \leq \left\vert
q\right\vert ^{sn},$ \textit{where} $s=\dim _{\mathbb{F}_{q}}R.$\medskip

\textbf{Proof}. We know that $R$ is a vector space over $\mathbb{F}_{q}$.
Let $s=\dim _{\mathbb{F}_{q}}R$ and $\{1,v_{1},...,v_{s-1}\}$ be a basis in\ 
$R.$ We will prove that\newline
$B$=$\{1,x$,...,$x^{n-1},v_{1},v_{1}x$,...,$v_{1}x^{n-1}$,...,$%
v_{s-1},v_{s-1}x$,...,$v_{s-1}x^{n-1}\}$ is a basis in the $\mathbb{F}_{q}$%
-vector space $R_{n},R_{n}\in \{R_{i,n},S_{i,n},T_{(i,j),n}\}.$

First, we will show the linearly independence of the elements from $B$. If
there are the elements $\alpha _{1,i_{1}},\alpha _{2i_{2}},...,\alpha
_{si_{s}}\in \mathbb{F}_{q}$ $i_{j}\in \{0,1,2,...,n-1\},j\in \{1,2,...,s\}$
such that $\alpha _{1,0}\cdot 1$+...+$\alpha _{1,n-1}x^{n-1}$+...+$\alpha
_{s,0}v_{s-1}$+...+$\alpha _{s,n-1}v_{s-1}x^{n-1}$=0, comparing the
coefficients in this equation, we get 
\begin{equation}
\alpha _{1,0}\cdot 1+\alpha _{2,0}v_{s-1}+...+\alpha _{s,0}v_{s-1}=0. 
\tag{3.1.}
\end{equation}

Since $v_{1},...,v_{s-1}$ are nilpotent elements in $R,$ if $\alpha
_{1,0}\neq 0,$ from relation $\left( 3.1\right) ,$ we obtain that a unit is
equal \ with a nilpotent element, false. Hence $\alpha _{1,0}=0$ and $\alpha
_{2,0}v_{s-1}+...+\alpha _{s,0}v_{s-1}=0,$ therefore $\alpha
_{2,0}=...=\alpha _{s,0}=0.$ In the same way, comparing coefficients of $%
x,x^{2},...,x^{n-1}$ with zero, we have $\alpha _{1,i_{1}}=\alpha
_{2,i_{2}}=...=\alpha _{s,i_{s}}=0,$ for all $i_{j}\in
\{0,1,2,...,n-1\},j\in \{1,2,...,s\}.$

We will prove that $B$ generates $R_{n}$. Let $f\left( x\right) \in R_{n}.~$%
By straightforward calculations, we obtain that $f\left( x\right) $ is a
linear combination of elements in $B$ with coefficients in $\mathbb{F}_{q}.$
It results $\left\vert R_{n}\right\vert =\left\vert q\right\vert ^{sn}.$

Now, let $C$ be a nonzero cyclic code. $C$ is a vector subspace of the $%
\mathbb{F}_{q}$-vector space $R_{n},~$therefore $\left\vert C\right\vert
\leq \left\vert q\right\vert ^{sn}.~\Box \medskip $

In [Ab, Si; 07], Theorem 3, and in [Al-As, Ha; 10], Theorem 4.2, the authors
gave a basis or a minimal spanning set for the codes of even length over $%
\mathbb{Z}_{2}+u\mathbb{Z}_{2},$ respectively $\mathbb{Z}_{2}+u\mathbb{Z}%
_{2}+...+u^{k-1}\mathbb{Z}_{2}.$ The same description could be done, in
general case, over the ring $R_{i}.\ $If we have supplementary relations
between polynomials $h_{1},...,h_{t},$ we can compute $\left\vert
C\right\vert ,$ as we can see in the following examples.\medskip

\textbf{Example 3.4.} \ Over $R_{i},$ for $i=2,$ using Theorem 3 from [Ab,
Si; 07], if $C$ is a nonzero cyclic code of length $n,\left( n,p\right) \neq
1,$ and $\ C=(g\left( x\right) +up\left( x\right) ,ua\left( x\right)
),a\left( x\right) \mid g\left( x\right) \mid \left( x^{n}-1\right) ,$ with $%
\deg g\left( x\right) =r,\deg a\left( x\right) =t,r\geq t,\deg a\left(
x\right) >\deg p\left( x\right) ,$ then $\left\vert C\right\vert
=(q)^{2n-r-t}.$ Indeed, $C$ is a vector space over $\mathbb{F}_{q}$ and let%
\newline
$B$=\{$g$($x$)+$up$($x$),$x$($g$($x$)+$up$($x$)),...,\newline
$x^{n-r-1}$($g$($x$)+$up$($x$)),$ua$($x$),$xua$($x$),...,$x^{n-t-1}ua$($x$)\}%
$.$ We will prove that $B$ is a basis in the $\mathbb{F}_{q}$-vector space $%
C.$

First, we will show that $B$ spans $C$. Let $c\left( x\right) \in C,$ then 
\newline
$c\left( x\right) =q_{1}\left( x\right) \left( g\left( x\right) +up\left(
x\right) \right) +q_{2}\left( x\right) ua\left( x\right) ,q_{i}\left(
x\right) \in R[x],i\in \{1,2\}.$ \newline
If$~\deg q_{1}\left( x\right) <n-r$ and $\deg q_{2}<n-t,$ we have $B$ spans $%
C$. If $\deg q_{1}\left( x\right) \geq n-r$ or $\deg q_{2}\geq n-t,$ it is
suffices to show that $x^{n-r}\left( g\left( x\right) +up\left( x\right)
\right) ,$\newline
$u\left( g\left( x\right) +up\left( x\right) \right) $ and $x^{n-t}ua\left(
x\right) $ are generated by $B$ over $\mathbb{F}_{q}.$

We have $x^{n-r}\left( g\left( x\right) +up\left( x\right) \right)
=x^{n}-1+q\left( x\right) ,\deg q\left( x\right) \leq n-1.$ But $q\left(
x\right) \in C$ and, by the division algorithm, we have $q\left( x\right)
=\left( g\left( x\right) +up\left( x\right) \right) h_{1}\left( x\right)
+s_{1}\left( x\right) ,$ $\deg s_{1}\left( x\right) <r,\deg h_{1}\left(
x\right) \leq n-1,$ $s_{1}\left( x\right) =ua\left( x\right) h_{2}\left(
x\right) +s_{2}\left( x\right) ,\deg s_{2}\left( x\right) <\deg ua\left(
x\right) ,\deg h_{2}\leq r-t.$ Since $\deg ua\left( x\right) =\deg a\left(
x\right) $ and in $C$ any polynomial must have degree greater or equal with $%
\deg a\left( x\right) ,$ it results $s_{2}\left( x\right) =0$. Since $%
a\left( x\right) \mid g\left( x\right) ,$ hence $g\left( x\right) =a\left(
x\right) h\left( x\right) $ and we have $ug\left( x\right) =u\left( g\left(
x\right) +up\left( x\right) \right) =ua\left( x\right) h\left( x\right) ,$
with $\deg h\left( x\right) \leq r-t<n-t-1.$ It results that $u\left(
g\left( x\right) +up\left( x\right) \right) $ is generated by $B$ over $%
\mathbb{F}_{q}.$ For finish the proof, it is enough to show that the element 
$ux^{r-t}a\left( x\right) $ is generated by $B$ over $\mathbb{F}_{q}.$ We
have $ux^{r-t}a\left( x\right) =u\left( g\left( x\right) +up\left( x\right)
\right) +uh_{3}\left( x\right) ,$ with $uh_{3}$ belongs to $C$ and $t\leq
\deg h_{3}\left( x\right) <r.$ Therefore $uh_{3}\left( x\right) =\alpha
_{0}ua\left( x\right) +\alpha _{1}xua\left( x\right) +$...\newline
$+\alpha _{r-t-1}x^{r-t-1}ua\left( x\right) ,\alpha _{i}\in \mathbb{F}%
_{q},i\in \{0,...,r-t-1\}.$

We will prove that $B$ is a linearly independent system. Indeed, if there
are the elements\newline
$\alpha _{1,i_{1}},\alpha _{2,i_{2}},\in \mathbb{F}_{q}$ $i_{j}\in
\{0,1,2,...,n-1\},j\in \{1,2\}$ such that \newline
$\alpha _{1,0}\left( g\left( x\right) +up\left( x\right) \right) $+...+%
\newline
$\alpha _{1,n-1}x^{n-r-1}\left( g\left( x\right) +up\left( x\right) \right) $%
+$\alpha _{2,0}ua\left( x\right) $+...+$\alpha _{2,n-1}x^{n-t-1}ua\left(
x\right) $=$0$, comparing the coefficients in this equation, we get \newline
$\alpha _{1,0}g\left( 0\right) +\alpha _{1,0}up\left( 0\right) +\alpha
_{2,0}ua\left( 0\right) =0.$ If \ $\alpha _{1,0}\neq 0,$ since $g\left(
0\right) $ is a unit and $u$ a nilpotent element, it results that a unit is
equal with a nilpotent, false. Therefore $\alpha _{1,0}=0.$ We obtain $%
\alpha _{2,0}ua\left( 0\right) =0.$ If $\alpha _{2,0}\neq 0,$ it results $%
ua\left( 0\right) =0,$ false, since $a\left( 0\right) $ is a unit. We repeat
this procedure and we get $\alpha _{1,i_{1}}=\alpha _{2,i_{2}}=0,$ for all $%
i_{j}\in \{0,1,2,...,n-1\},j\in \{1,2\},$ hence $\ B$ is a linearly
independent system. It results that $B$ is a basis in the $\mathbb{F}_{q}$%
-vector space $C$ and $\left\vert C\right\vert =\left\vert q\right\vert
^{2n-r-t}.~\Box $

\begin{equation*}
\end{equation*}%
\begin{equation*}
\end{equation*}%
\begin{equation*}
\end{equation*}
\begin{equation*}
\end{equation*}
\begin{equation*}
\end{equation*}

{\large 4.} {\large Minimum Hamming distance for the cyclic codes over the
rings \ }$R_{i},S_{i},T_{(i,j)}\bigskip \bigskip $

Let $C$ be a linear code over the ring $R.$ The \textit{Hamming distance}
between two codewords $c_{1}$ and $c_{2},$ denoted $H\left(
c_{1},c_{2}\right) ,$ is the number of coordinates in which the codewords $%
c_{1}$ and $c_{2}$ differ. \ The number of nonzero entries of a codeword $c$%
, denoted $w\left( c\right) $, is called the \textit{Hamming weight} of the
codeword $c.$ The \textit{Hamming distance} of a linear code $C$ is 
\begin{equation*}
d\left( C\right) =\text{\textit{min}}\{w\left( c\right) \text{ }/~c\in
C,c\neq 0\}.
\end{equation*}%
In [Ab, Si; 06], the authors studied the Hamming distance of cyclic codes of
even length, especially codes of length $2^{e},e\in \mathbb{N}-\{0\}$.
(Lemma 16, Lemma 17, Lemma 18). In the following, using some ideas from the
mentioned lemmas,we will investigate the Hamming distance for cyclic codes
of length $n=p^{r},r\in \mathbb{N}-\{0\},$ over the rings $%
R_{i},S_{i},T_{(i,j)}.\medskip $

\textbf{Definition 4.1.} Let $%
n=a_{s-1}p^{s-1}+a_{s-2}p^{s-2}+...+a_{1}p^{1}+a_{0}p^{0},$ $\alpha _{i}\in
\{0,1,...,p-1\},i\in \{0,1,...,s-1\},$ be the $p$-adic expansion of $n.$

1) If $a_{s-1}=...=a_{s-t}\neq 0,s-t>0$ and $a_{s-i}=0,$ for all $i\in
\{t+2,t+3,...,s-1\},$ then $n$ has a $p$-adic length $t$ zero expansion

2) If $a_{s-1}=...=a_{s-t}\neq 0,s-t>0$ and $a_{s-i}\neq 0,$ for some
elements $i\in \{t+2,t+3,...,s-1\}$, then $n$ has a $p$-adic length $t$
non-zero expansion.

3) If $s=t,$then $n$ has a $p$-adic full expansion.\medskip

\textbf{Proposition 4.2.} \textit{Let} $C=(g\left( x\right) )$ \textit{be a
cyclic code over }$R\in \{R_{i},S_{i},T_{(i,j)}\}$\textit{\ of length} $%
p^{r},r\in \mathbb{N}-\{0\},$ \textit{where} $g\left( x\right) =\left(
x^{ap^{r-1}}-1\right) g_{1}\left( x\right) .$ \textit{If} $\ g_{1}\left(
x\right) $ \textit{generates a cyclic code of length} $p^{r-1}$ \textit{and
Hamming distance} $d,$ \textit{then} $d\left( C\right) =2d.\medskip $

\textbf{Proof.} \ For $c\in C$ we have $c=\left( x^{ap^{r-1}}-1\right)
g_{1}\left( x\right) g_{2}\left( x\right) ,$\newline
$g_{2}\left( x\right) \in \mathbb{F}_{q}[x]/(x^{n}-1)$ and $g_{1}\left(
x\right) g_{2}\left( x\right) \in \left( g_{1}\left( x\right) \right) .$ 
\newline
It results $\ w\left( c\right) =w\left( \left( x^{ap^{r-1}}-1\right)
g_{1}\left( x\right) g_{2}\left( x\right) \right) =w\left(
x^{ap^{r-1}}g_{1}\left( x\right) g_{2}\left( x\right) \right) +w\left(
g_{1}\left( x\right) g_{2}\left( x\right) \right) .$ Then $d\left( C\right)
=d+d=2d.~\Box \medskip $

\begin{equation*}
\end{equation*}

\textbf{Conclusions.} In this paper we investigate the structure of cyclic
codes of arbitrary length over the rings $R_{i},S_{i},T_{(i,j)}.$ Moreover
the ranks and minimum Hamming distance of these codes was studied. Since the
rings with Hamming weight cannot produce always better codes, a more
relevant weight as, for example, the homogeneous weight on the above
mentioned rings can be studied. The remark above can constitute the starting
point for further research.\bigskip

\begin{equation*}
\end{equation*}%
\textbf{Acknowledgements.} I would like to thank the referee for his/her
many suggestions which helped me improve this paper.%
\begin{equation*}
\end{equation*}

\textbf{References}

\begin{equation*}
\end{equation*}

[Ab, Si; 07] Abualrub, T., Siap, I., \textit{Cyclic codes over the rings }%
\textbf{\ }$\mathbb{Z}_{2}+u\mathbb{Z}_{2}$ \textit{and} $\mathbb{Z}_{2}+u%
\mathbb{Z}_{2}+u^{2}\mathbb{Z}_{2},$ Des Codes Cryptogr., \textbf{42}(2007),
273-287.

[Ab, Si; 06] Abualrub, T., Siap, I., \textit{On the construction of cyclic
codes over the ring }\textbf{\ }$\mathbb{Z}_{2}+u\mathbb{Z}_{2},$
Proceedings of the 9th WSEAS International Conference on Applied
Mathematics, Istanbul, Turkey, May 27-29, 2006, 430-435.

[Al-As, Ha; 10] Al-Ashker, M.M., Hamoudeh, M., \textit{Cyclic codes over} $%
\mathbb{Z}_{2}+u\mathbb{Z}_{2}+...+u^{k-1}\mathbb{Z}_{2},$ Turk J Math, 
\textbf{34}(2010), 1-13.

[Bh, Wa; 09] Bhaintwal, M., Wasan, S.,K., \textit{On quasi-cyclic codes over 
}$\mathbb{Z}_{q},$ AAECC, \textbf{20}(2009), 459-480.

[Bl; 72] Blake, I.F., \textit{Codes over certain rings}, Inf. Control, 
\textbf{20}(1972), 396-404.

[Do, Li, Pa; 11] \ Dougherty, S. T., Liu, H., Park, Y. H., \textit{Lifted
codes over finite chain rings}, Mathematical Journal of Okayama University, 
\textbf{53}(2011), 39--53.

[Do, Ka, Yi; 11] Dougherty, S. T., Karadeniz, S., Yildiz, B., \textit{Cyclic
codes over} $R_{k}$, Des Codes Cryptogr., DOI 10.1007/s10623-011-9539-4.

[Ei; 95] Eisenbud, D., \textit{Commutative algebra}, Graduate Texts in
Mathematics, \textbf{150}, Springer-Verlag, Berlin, New York, 1995.

[Gr; 97] Greferath, M., \textit{Cyclic codes over \ finite rings,} Discrete
Math., \textbf{177(3)}(1997), 273-277.

[Ha et all; 94] Hammons, A.R. Jr., Kumar, P.V., Calderbank, A.R., Sloane,
N.J.A., Sol\'{e}, P., \textit{The }$Z_{4}$\textit{\ liniarity of Kerdock,
Preparata, Goethals and related codes}, IEEE Trans. Inf. Theory, \textbf{40}%
(1994), 301-319.

[Hu; 74], Hungerford, T.W.,\textit{\ Algebra}, Springer Verlag, New York,
1974.

[McD; 74] McDonald, B.R., \textit{Finite Rings with Identity}, New York:
Marcel Dekker Inc., \ 1974.

[Mi; 80], Milne, J. G., \textit{\'{E}tale cohomology}, Princeton University
Press, 1980.

[Ne, Ho; 99] Nechaev, A. A., Honold, T., \textit{Fully weighted modules and
representations of codes}, (Russian) Problemy Peredachi Informatsii \textbf{%
35(3)}(1999), 18-39; translation in Problems Inform.Transmission \textbf{35}%
(3)(1999), 205-223.

[Qi, Zh, Zhu; 05] Qian, J.-F., Zhang, L.-N., Zhu, A.-X., \textit{Cyclic
codes over} $\mathbb{F}_{p}+u\mathbb{F}_{p}+...+u^{k-1}\mathbb{F}_{p},$
IEICE Trans. Fundamentals, Vol. E88-A, \textbf{3}(2005), 795-797.

[So, Si; 07] Sol\'{e}, P., \ Sison, V., \textit{Bounds on the Minimum
Homogeneous Distance of the} $p^{r}-$ \textit{ary Image of Linear Block
Codes Over the Galois Ring GR}$\left( p^{r},m\right) ,$ IEEE Transactions on
Information Theory , \textbf{53(6)}(2007), 2270-2273.

[So, Si; 07(1)] Sol\'{e}, P., \ Sison, V., Quaternary Convolutional Codes
From Linear Block Codes Over Galois Rings , IEEE Transactions on Information
Theory , \textbf{53(6)}(2007), 2267-2270.

[Sp; 78] Spiegel, E., \textit{Codes over} $Z_{m}\ \ $\textit{revisited},
Inform. and Control, \textbf{37}(1978), 100--104.

[Yi, Ka; 11] Yildiz, B., Karadeniz, S., \textit{Cyclic codes over} $\mathbb{F%
}_{2}+u\mathbb{F}_{2}+v\mathbb{F}_{2}+uv\mathbb{F}_{2},$ Des. Codes
Cryptogr., \textbf{58}(2011), 221-234.

\end{document}